\documentclass[aps,twocolumn,nofootinbib,preprintnumbers,prl,10pt]{revtex4-2}
\usepackage[utf8]{inputenc}
\usepackage[svgnames]{xcolor}
\usepackage[most]{tcolorbox}
\usepackage{caption}
\tcbset{colback=PaleGreen!0!white, colframe=NavyBlue, 
	highlight math style= {enhanced, 
		colframe=red,colback=red!10!white,boxsep=0pt}
}
\usepackage{relsize}

\usepackage{amsmath}
\usepackage[bottom]{footmisc}
\usepackage{braket}
\usepackage{graphicx}
\usepackage{mwe}
\usepackage[section]{placeins}
\usepackage{framed}
\usepackage{csquotes}
\usepackage{tikz}
\usetikzlibrary{decorations.markings}
\usetikzlibrary{decorations.pathmorphing}
\definecolor{shadecolor}{rgb}{0.90,0.90,0.90}
\usepackage{hyperref}
\usepackage{bbold}
\usepackage{subcaption}
\usepackage{pifont}
\usepackage{setspace}
\usepackage{amsmath, amssymb, amsthm, float, graphicx,amsfonts}
\usepackage{amsthm}
\usepackage{array, boldline, makecell, booktabs}

\theoremstyle{definition}

\hypersetup{colorlinks=true, linkcolor=DarkRed, citecolor=DarkRed,urlcolor=Indigo,linktocpage}
\def\beq{\begin{eqnarray}}\def\eeq{\end{eqnarray}}
\def\be{\begin{equation}}\def\ee{\end{equation}}
\def\bs{\begin{split}}\def\es{\end{split}}

\def\r{\rho}
\def\s{\sigma}
\def\m{\mu}

\def\a{\alpha}

\def\d{\delta}

\def\mW{{\mathcal{W}}}

\def\oforder#1{\mathinner{O\left({#1}\right)}}
\usepackage{bm}

\begin{document}
\title{\bf Crossing Symmetric Dispersion Relations  in QFTs}
\author{Aninda Sinha$^{a}$\footnote{asinha@iisc.ac.in}  and Ahmadullah Zahed$^{a}$\footnote{ahmadullah@iisc.ac.in}\\
\it ${^a}$Centre for High Energy Physics,
\it Indian Institute of Science,\\ \it C.V. Raman Avenue, Bangalore 560012, India. }

\begin{abstract}{For 2-2 scattering in quantum field theories, the usual fixed $t$ dispersion relation exhibits only two-channel symmetry. This paper considers a crossing symmetric dispersion relation, reviving certain old ideas in the 1970s. Rather than the fixed $t$ dispersion relation, this needs a dispersion relation in a different variable $z$, which is related to the Mandelstam invariants $s,t,u$ via a parametric cubic relation making the crossing symmetry in the complex $z$ plane a geometric rotation. The resulting dispersion is manifestly three-channel crossing symmetric. We give simple derivations of certain known positivity conditions for effective field theories, including the null constraints, which lead to two sided bounds and derive a general set of new non-perturbative inequalities. We show how these inequalities enable us to locate the first massive string state from a low energy expansion of the four dilaton amplitude in type II string theory. We also show how a generalized (numerical) Froissart bound, valid for all energies, is obtained from this approach.}
\end{abstract}
\maketitle

{\bf Introduction:} Dispersion relations provide non-perturbative representations for scattering amplitudes in quantum field theories \cite{mandelstam, nuss}. The usual way to write dispersion relations in the context of 2-2 scattering of identical particles is to keep one of the Mandelstam invariants, usually $t$, fixed and write a complex integral in the variable $s$. This approach naturally leads to an $s-u$ symmetric representation of the amplitude. Then, one imposes crossing symmetry as an additional condition. A similar approach can also be developed for Mellin amplitudes for conformal field theories. Recent developments in this direction include \cite{joaopaper}, \cite{joaopaper2}, \cite{cmrs}.

The amplitude's resulting representation not having manifest three-channel crossing symmetry may appear to be a drawback. For instance, in perturbative quantum field theories, when we compute Feynman diagrams, the amplitude's resulting expansion exhibits crossing symmetry. In the worldsheet formulation of string theory, the tree level sphere diagram, for instance, is also manifestly crossing symmetric. Hence, it seems like a natural question to ask as to how would one directly see the structure of Feynman diagrams from dispersion relations.

 Is there a crossing symmetric version of the dispersion relations? In the 1970s, this question was briefly considered in a few papers, for example, in 1972 by Auberson and Khuri in \cite{AK} and in 1974 by Mahoux, Roy, and Wanders in \cite{mahoux}. Unfortunately, due to the technical complications involved, barring for a smattering of a few papers (e.g., \cite{anant}), this approach has not been well explored in the literature. We will follow \cite{AK} and revive this line of questioning again. In the CFT context, Polyakov's work in \cite{Pol} proposed a fully crossing symmetric bootstrap, which was developed in \cite{ks, usprl}. However, this approach currently lacks a non-perturbative derivation for $d\geq 2$. Our methods in this paper will enable us to address this important question in the near future \cite{GSZ}.

Dispersion relations also give a window to understanding how analyticity and unitarity assumptions for the high energy behaviour of amplitudes constrain low energy physics contained in effective field theories (EFTs) \cite{nima}.
Our manifestly crossing symmetric approach not only leads to a simpler and unifying derivation of recently considered positivity constraints in EFTs \cite{RMTZ, rattazzi, TWZ, Caron-Huot:2020cmc}, but also enables us to write down a completely general set of positivity constraints on the Wilson coefficients. In particular, we will provide straightforward derivations of many of the upper bounds on the ratios of Wilson coefficients, as well as the null constraints listed in \cite{TWZ, Caron-Huot:2020cmc}, leading to the lower bounds. Our formalism will enable us to write down general formulae for the upper bounds and the independent null constraints.

We will consider two novel applications of our constraints. First, using them, we will locate the first massive string pole from the low energy expansion for the tree-level four-dilaton scattering in type II string theory (eg.\cite{green}). Second, our approach enables us to derive a numerical upper bound on the total scattering cross-section of identical particles valid at all energies, generalizing the famous Froissart bound \cite{froissart}. Further, we will explain how our approach leads to a structure like the Feynman diagram expansion in QFTs.\\
{\bf Crossing symmetric dispersion relation: }
We begin by considering cubic hypersurfaces  \cite{AK} in the variables
$
s_1=s-\frac{\m}{3},~s_2=t-\frac{\m}{3},~s_3=u-\frac{\m}{3}=-s_1-s_2,~ \m=4m^2$, where $s,t,u$ are usual Mandelstam variables. Explicitly, these hypersurfaces will be given by $
\left(s_{1}(z)-a\right)\left(s_{2}(z)-a\right)\left(s_{3}(z)-a\right)=-a^3,$ with $a$ being a real parameter \cite{foot1}.  The $s_i$'s can be parametrized via
\be
\label{eq:skdef}
s_{k}=a- \frac{a \left(z-z_{k}\right)^{3}}{z^{3}-1}, \quad k=1,2,3\,,
\ee
where $z_k$ are cube roots of unity and we will restrict $-\frac{\m}{3}\leq a<\frac{2\m}{3}$. In \cite{AK}, it is shown that the amplitude is analytic in the interval $-6.71\mu <a<2\mu/3 $. Importantly, note that $a=\frac{s_1 s_2 s_3}{s_1s_2+s_2s_3+s_3s_1}$ is crossing symmetric in the usual variables. The amplitude $\mathcal{M}(s_1,s_2)$ can be written as an analytic function of $(z,a)$ \textit{i.e.,} $\overline{\mathcal{M}}(z,a)=\mathcal{M}\left(s_1(z),s_2(z)\right)$. $\mathcal{M}\left(s_1(z),s_2(z)\right)$ has physical cuts for $s_k\geq \frac{2\m}{3}, ~k=1,2,3$. These physical cuts get mapped to portions of arcs on the unit circle in the complex $z$-plane.
We can write down a twice subtracted dispersion relation  $z$-variable, for fixed $a$. For the completely crossing symmetric case, the dispersion relation simplifies dramatically in terms of  the $s_1,s_2,s_3$ variables:
\be
\label{eq:disper00}
\begin{split}
\mathcal{M}_{0}(s_1, s_2)=\alpha_{0}+\frac{1}{\pi} \int_{\frac{2\m}{3} }^{\infty}& \frac{d s_1^{\prime}}{s_1^{\prime}} \mathcal{A}\left(s_1^{\prime} ; s_2^{(+)}\left(s_1^{\prime} ,a\right)\right)\\
&\times  H\left(s_1^{\prime} ;s_1, s_2, s_3\right)\,,
\end{split}
\ee
where $\mathcal{A}\left(s_1; s_2\right)$ is the s-channel discontinuity and
\be\nonumber
\begin{split}
&H\left(s_1^{\prime} ; s_1, s_2,s_3\right)=\left[\frac{s_1}{\left(s_1^{\prime}-s_1\right)}+\frac{s_2}{\left(s_1^{\prime}-s_2\right)}+\frac{s_3}{\left(s_1^{\prime}-s_3\right)}\right]\\
&s_{2}^{(+)}\left(s_1^{\prime}, a\right)=-\frac{s_1^{\prime}}{2}\left[1 - \left(\frac{s_1^{\prime}+3 a}{s_1^{\prime}-a}\right)^{1 / 2}\right]\,,
\end{split}
\ee
which defines the crossing symmetric kernel $H$ \cite{supp1}. $\alpha_0$ is a subtraction constant. From hereon, we will follow a more direct route than what was followed in \cite{AK}. This final form in eq.(\ref{eq:disper00}) is manifestly three channel crossing symmetric. Notice the nontrivial $s_2^{(+)}$ dependence in $\mathcal{A}$--we emphasise that the crossing symmetric $s_1,s_2,s_3$ dependence also comes via $s_2^{(+)}$ through $a$. One can use the crossing symmetric form and carry out several checks. For instance, we have numerically investigated how well the representation represents the type II superstring four dilaton amplitude \cite{supp1}. For this case, on the real $(s_1,s_2)$ plane, the crossing symmetric formula is often a better representation than the conventional dispersion relation. \\
%
{$~~~~~$}For the situation where we have no massless poles, or where we subtract them out, the crossing symmetric amplitude has an expansion 
\be
\label{eq:cpqdef}
\mathcal{M}_{0}(s_1,s_2)=\sum_{p, q=0}^{\infty} {\mathcal W}_{p q} x^{p} y^{q}\,,
\ee
with $x=-\left(s_1 s_2 + s_2 s_3+s_3 s_1\right)$, $y=-s_1 s_2 s_3$. Note that $H\left(s_1^{\prime} ; s_1, s_2,s_3\right)=\frac{ x \left(2 s_1'-3a\right)}{x a -x s_1'+s_1^{'3}}\,,$ which can be seen by writing $s_k$'s in terms of $(z,a)$ and identifying $\frac{z^{3}}{\left(z^{3}-1\right)^{2}}=\frac{-x}{27a^2}$. We can now expand \eqref{eq:disper00} in powers of $x, a$
%
using the s-channel discontinuity which has a partial wave expansion in terms of Gegenbauer polynomials involving {\it even} spins 
{\begin{small}
\be\nonumber
\begin{split}
&\mathcal{A}\left(s_1,s_2^{(+)}(s_1,a)\right)=\Phi(s_1)\sum_{\ell=0}^{\infty}\left(2\ell+2\a\right)a_\ell(s_1)C^{(\a)}_{\ell}\left(\sqrt{\xi(s_1,a)}\right)\,,\\
&\xi(s_1,a)=\xi_0+4\xi_0\left(\frac{a}{s_1-a} \right),~\xi_0=\frac{s_1^2}{(s_1-2\mu/3)^2}
\end{split}
\ee
\end{small}}\\
where $\a=\frac{d-3}{2}$, $\Phi(s_1)=\Psi(\alpha)\frac{\sqrt{s_1+\frac{\mu}{3}}}{\left(s_1-\frac{2\mu}{3}\right)^{\a}}$ and $\Psi(\a)$ is  real positive number. Expanding the Gegenbauer polynomials around $\xi=\xi_{0}$, 
with $p_{\ell}^{(j)}\left(\xi_{0}\right)=\partial^j C^{(\a)}_{\ell}\left(\sqrt{\xi}\right)/{\partial{\xi^j}}{ |}_{\xi=\xi_0},
$ we find the coefficient of $a^m$ leading to an inversion formula
\be
\begin{split}\label{eq:cpqallform}
&\mW_{n-m,m}=\int_{\frac{2\m}{3}}^{\infty}\frac{d s_1}{s_1}\Phi(s_1)\sum_{\ell=0}^{\infty}\left(2\ell+2\a\right)a_\ell(s_1)\mathcal{B}_{n,m}^{(\ell)}(s_1)\,,\\
&\mathcal{B}_{n,m}^{(\ell)}(s_1)=\sum_{j=0}^{m}\frac{p_{\ell}^{(j)}\left(\xi_{0}\right) \left(4 \xi_{0}\right)^{j}(3 j-m-2n)(-n)_m}{\pi s_1^{2 n+m} j!(m-j)!(-n)_{j+1}}\,,
\end{split}
\ee
for $~n\geq 1$. Note that $\mW_{0,0}=\alpha_0$.
Eq.\eqref{eq:cpqallform} allows two lines of investigation. {\bf (a)} For $n\geq m$ with $n\geq 1$, we get the coefficients in terms of $a_\ell(s_1)$. Since partial wave unitarity implies $0\leq a_\ell(s_1)\leq 1$, we can find positivity constraints on $\mW_{n-m,m}$. {\bf (b)} For $n<m$ with $n\geq 1$, the coefficients should vanish, as needed by eq.\eqref{eq:cpqdef}, which give rise to  non-trivial constraints on $a_\ell(s_1)$. Notice that since $\ell$ is even, we will need $\ell\geq 2n$. These sum rules or ``null constraints" are instrumental in getting the lower bounds on $\mW_{n-m,m}$ in EFTs. We will use these sum rules to put a bound on the total cross section at any $s_1$, generalizing the famous Froissart bound. \\
{\bf Constraining QFTs}:
Now from eq \eqref{eq:cpqallform}, we can derive inequalities involving $\mW_{p,q}$. 
We use the unitarity constraints $0\leq a_{\ell}(s_1)\leq 1$ as well as 
$C_{\ell-k }^{(\alpha+k)}\left(\frac{2 \mu }{3 \delta }+1\right)\geq 0,$ for $\d=s_1-\frac{2\m}{3}\geq 0, ~\a\geq 0,~\m\geq0 \,$. Since the range of $s_1$ in eq.(\ref{eq:cpqallform}) starts at $2\mu/3$, we have introduced $\delta$ as a convenient variable.  We have $\sqrt{\xi_0}=\frac{2 \mu }{3 \delta }+1$.
Note that $p_\ell^{(j)}$'s involve derivatives of Gegenbauer. Specifically we find the useful expression
\begin{small}
\be\nonumber
\mathcal{B}^{(\ell)}_{n,m}(s_1)=\sum_{k=0}^{m}\mathfrak{U}^{(\a)}_{n,m,k} (-1)^{k+m}C_{\ell-k }^{(\alpha+k)}\left(\frac{2 \mu }{3 \delta }+1\right)\,,
\ee
\end{small}
where \cite{foot3}
\begin{small}
\be\nonumber
\label{eq:Bellnmk}
\mathfrak{U}^{(\a)}_{n,m,k}=\sum _{j=k}^m \frac{\sqrt{16 \xi _0}^k (\alpha )_k (m+2 n-3j) \Gamma (n-j) \Gamma (2 j-k)}{s_1^{m+2 n}\Gamma (k) j! (m-j)!(j-k)!(n-m)!}\,,
\ee
\end{small}
is positive for $n\geq m$. Note that in the sum,  $(-1)^{k+m}$ spoils the definite sign of $\mathcal{B}^{(\ell)}_{n,m}(s_1)$. We can search for $\chi _n ^{(r,m)}(\m,\d)$, such that 
\be\nonumber
\label{eq:Bellnmall}
\begin{split}
&\sum_{r=0}^{m}\chi _n ^{(r,m)}(\m,\d)\mathcal{B}^{(\ell)}_{n,r}(s_1)=\mathfrak{U}^{(\a)}_{n,m,m} C_{\ell-m }^{(\alpha+m)}\left(\frac{2 \mu }{3 \delta }+1\right)\geq 0,
\end{split}
\ee
for $m\leq n$. 
A solution is easily found using the following recursion relation:
\be\label{rec}
\begin{split}
&\chi_n^{(r,m)}(\mu,\d)=\sum_{j=r+1}^m (-1)^{j+r+1}\chi_n^{(j,m)} \frac{\mathfrak{U}^{(\a)}_{n,j,r}(\frac{2\mu}{3}+\d)}{\mathfrak{U}^{(\a)}_{n,r,r}(\frac{2\mu}{3}+\d)}\,, \nonumber
\end{split}
\ee
with $\chi _n ^{(m,m)}(\m,\d)=1$. Using the recursion relation, one can check that \cite{exp1} $\chi_n^{(r,m)}>0$, for $n\geq m$. Then eq.(\ref{eq:cpqallform}) leads to \cite{comm1}
\be\label{eq:nonpcpq1}
\begin{split}
\sum_{r=0}^{m}\chi _n ^{(r,m)}(\m,\d=0)\mW_{n-r,r}\geq 0,
\end{split}
\ee 
which we will refer to as non-perturbative constraints to differentiate from the EFT constraints to be derived next.
In order to derive EFT bounds, we start with eq.\eqref{eq:cpqallform} and write
\begin{small}
\be\label{Weff}
\mW^{(\d_0)}_{n-m,m}\equiv\int_{\d_0+\frac{2\m}{3}}^{\infty} \frac{d s_1}{s_1}\Phi(s_1)\sum_{\ell=0}^{\infty}\left(2\ell+2\a\right)a_\ell(s_1)\mathcal{B}_{n,m}^{(\ell)}(s_1)\,,
\ee
\end{small}
for $~n\geq 1$, which defines for us the Wilson coefficients and are the $\mW$'s in eq.\eqref{eq:cpqallform} when $\d_0=0$. 
In such cases, we can show that
\be\label{eq:Bellnmall2}
\begin{split}
\sum_{r=0}^{m}\chi _n ^{(r,m)}(\m,\d_0)\mathcal{B}^{(\ell)}_{n,r}\left(\d+\frac{2\mu}{3}\right)\geq 0,
\end{split}
\ee
for $\d\geq\d_0$. 
This leads to positivity constraints  
\be\label{eq:nonpcpq}
\begin{split}
\sum_{r=0}^{m}\chi _n ^{(r,m)}\left(\m,\d=\d_0\right)\mW^{(\d_0)}_{n-r,r}\geq 0,
\end{split}
\ee
for $~\m\geq0,~\d_0\geq 0, ~m\leq n$. 
{Since $\mathcal{B}^{(\ell)}_{n,0}\left(\d+\frac{2\m}{3}\right)=\frac{2  C_{\ell }^{(\alpha )}\left(\frac{2 \mu }{3 \delta }+1\right)}{\pi \left(\delta +\frac{2 \mu }{3}\right)^{2 n} }$, we have
\be\label{eq:Wn0only}
\mW_{n,0}^{(\d_0)}\leq \frac{1}{\left(\d_0+\frac{2\m}{3}\right)^2}\mW_{n-1,0}^{(\d_0)}\,.
\ee
Our expressions are in agreement with the limited number of cases known in the literature \cite{RMTZ, TWZ,Caron-Huot:2020cmc} except that our derivation is manifestly crossing symmetric from the start and admit a straightforward generalization \cite{known}.

An immediate application of our general formulae is the examination of the $n\gg m$ limit. We find simply
\be
\sum _{r=0}^m ~\frac{n^{m-r}}{ (m-r)!~ \left(\delta _0+\frac{2 \mu }{3}\right)^{m-r}}~\frac{\mW^{(\d_0)}_{n-r,r}}{\mW^{(\d_0)}_{n,0}}\geq 0\,,
\ee
for $n\gg m$. We have checked that tree level type II string theory, to be discussed below, respects this.\\
{\bf Null constraints: }
To derive lower bounds, we make use of the $n<m$ vanishing conditions arising from eq.(\ref{eq:cpqallform}). In the large $\d$ limit, we have \cite{foot4}
\be\nonumber
\begin{split}
&\mathcal{B}_{n,m}^{(\ell)}(\d)=\frac{C_{\ell }^{(\alpha )}(1)}{\pi}~\frac{D^{(n,m)}_{\ell,\a}}{\d^{2n+m}}+\oforder{\frac{\m}{\d^{2n+m+1}}}\,,\\
&D^{(n,m)}_{\ell,\a}=\sum_{j=0}^{m}\frac{(-4)^j \left(-\frac{\ell }{2}\right)_j \left(\alpha+\frac{\ell }{2}\right)_j}{\left(\alpha +\frac{1}{2}\right)_j} \frac{(3j-m-2n) \Gamma(m-n)}{j! (m-j)! (j-n)!}\,.
\end{split}
\ee
Then in the limit when $\d_0\gg \mu$ we have \cite{foot5}
\begin{small}
\be\label{eq:largesDell}
\int_{\d_0}^{\infty} \frac{d s_1}{s_1^{\alpha+1/2}} ~\sum_{\ell=2}^{\infty}\left(2\ell+2\a\right)~a_\ell(s_1)~\frac{C_{\ell }^{(\alpha )}(1)}{\pi}~\frac{D^{(n,m)}_{\ell,\a}}{s_1^{2n+m}}=0\,.
\ee
\end{small}
{for $m>n,~n\geq 1$.}
For example $m=2,n=1$ gives 
$
D^{(1,2)}_{\ell,\a}=\frac{2 \ell  (\ell+2 \alpha ) (-11-10 \alpha +2 \ell  (\ell+2 \alpha  ))}{  (2 \alpha +1) (2 \alpha +3) }\,,
$ 
which was first derived in \cite{TWZ} from fixed-$t$ dispersion relations \cite{foot6}. 
Once these null constraints are in place, a judicious use of Cauchy-Schwarz inequality as used in \cite{TWZ}, or a more constraining numerical argument used in \cite{Caron-Huot:2020cmc} can be pursued to derive lower bounds. The existence of such bounds was originally emphasized in \cite{nimayutin}. Our approach gives completely general expressions for the independent null constraints.\\
{\bf Applications: } 
We will now consider two applications. The first application will make use of the $n\geq m$ constraints while the second will use the $n<m$ constraints arising from eq.(\ref{eq:cpqallform}).

{\it Tree level type II superstring theory:}  
The four dilaton type II superstring tree amplitude is given by \cite{green}
$$
\mathcal{M}(s_1,s_2)=-\frac{\Gamma \left(-s_1\right) \Gamma \left(-s_2\right) \Gamma \left(s_1+s_2\right)}{\Gamma \left(s_1+1\right) \Gamma \left(-s_1-s_2+1\right) \Gamma \left(s_2+1\right)}\,.
$$
We consider the amplitude $\mathcal{M}^{(cl)}$ obtained after subtracting out the massless pole $-\frac{1}{s_1 s_2 (s_1+s_2)}$.
We can easily compute the $\mW_{p,q}$, for example \cite{green},
$
\mW_{0,0}= 2 \zeta (3),~\mW_{0,1}= -2 \zeta (3)^2,~\mW_{0,2}= \frac{2}{3} \left(2 \zeta (3)^3+\zeta (9)\right)\,,
$ 
$\mW_{1,0}= 2 \zeta (5),~\mW_{1,1}= -4 \zeta (3) \zeta (5)$. Suppose we are given the first few terms in the derivative expansion. Then, can we say where the first massive string pole would occur? More precisely in eq.(\ref{Weff}), what is the maximum $\d_0$ we can use? Using the methods in this paper, we can address this question.

From $m=1,n=1$ (6 derivatives) condition  \eqref{eq:nonpcpq} ($\m=0$) gives us $ \frac{3 \zeta (5)}{\delta _0}-2 \zeta (3)^2>0$, which implies $\d_0<1.07644$. Similarly from other conditions, we can show that $\d_0<\d_0^{(\max)}$.
\begin{figure}[hbt!]
    \centering\includegraphics[width=0.42\textwidth]{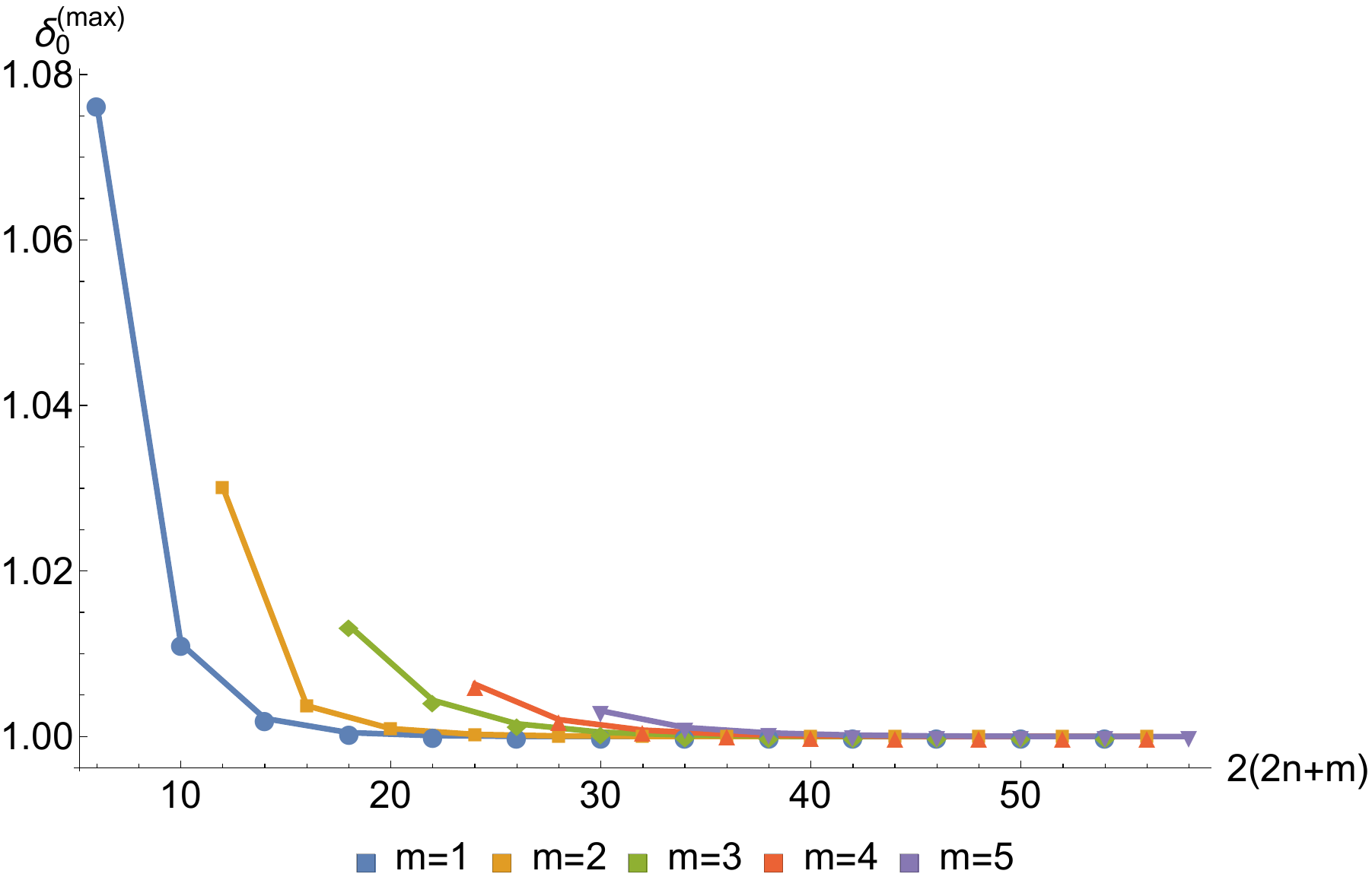}
  \caption{$\d_0^{(\max)}$ vs $2(2n+m)$, the derivative order, for $m=1,2,3,4,5$. }
  \label{fig:type2string}
\end{figure}
In figure \eqref{fig:type2string}, we have shown that for higher constraints \cite{foot7}, the $\d_0^{(\max)}$ converges towards $1$, which is exactly the location of the first massive pole!


{\it Bound on total scattering cross-section:}
Now, we will exploit the null constraints arising from $m>n$ in \eqref{eq:cpqallform} to bound total scattering cross-sections. We will be in $d=4$ or $\alpha=1/2$ and we will use the standard notation $s=s_1+4/3$ with $\m=4$. The null constraints read
\be\label{sum00}
\int_{4}^{\infty}\frac{ds}{s-\frac{4}{3}} \Phi(s) \sum_{\ell=2 n}^{\infty}(2 \ell+1) a_{\ell}(s) {\mathcal B}_{m,n}^{(\ell)}(s-\frac{4}{3})=0,
\ee
where $\Phi(s)=\sqrt{\frac{s}{s-4}}$.
for $m>n ,~ n \geq 1$. 
For $m+n\leq \ell,~m>n,~n\geq 1$ we can verify that ${\mathcal B}_{m,n}^{(\ell)}(s-\frac{4}{3})\geq 0$. Then we can write \eqref{sum00} as
\be\label{sum01}
\begin{split}
\int_{4}^{\infty}\frac{ds}{s-\frac{4}{3}} \Phi(s)\sum_{\ell=2 n}^{L_{max}} (2\ell+1)a_{\ell}(s)  {\mathcal B}_{m,n}^{(\ell)}(s-\frac{4}{3})\leq 0 \,,
\end{split}
\ee
for $ m+n \leq L_
{max},~ m>n,~n\geq 1$. 
We have placed the contributions arising from $\ell\geq L_{max}+2$ on the right which gives the inequality.
From unitarity, we know that $0 \leq a_{\ell}(s)\leq 1$. The inequalities \eqref{sum01} impose further conditions on the $a_{\ell}(s)$. We convert the integral over $s$ in \eqref{sum01} as a sum by defining $s(k)=4+k \frac{(s_{max}-4)}{N_{max}}$.
Using the constraints \eqref{sum01}, we want to bound the total scattering cross section $\s(s)$ \cite{foot8}
\be
\s(s)=\frac{16 \pi}{s-4}\sum_{\ell=0}^{\infty}(2\ell+1)a_{\ell}(s)\approx \frac{16 \pi}{s-4}\sum_{\ell=0}^{L_{max}}(2\ell+1)a_{\ell}(s) \,.
\ee
We maximize the value of  
$
\frac{s-4}{16\pi} \times \s(s)=\bar{\s}= \sum_{\ell=0}^{L_{max}}(2\ell+1)a_{\ell}(s)
$. 
\begin{figure}[hbt!]
    \centering\includegraphics[width=0.4\textwidth]{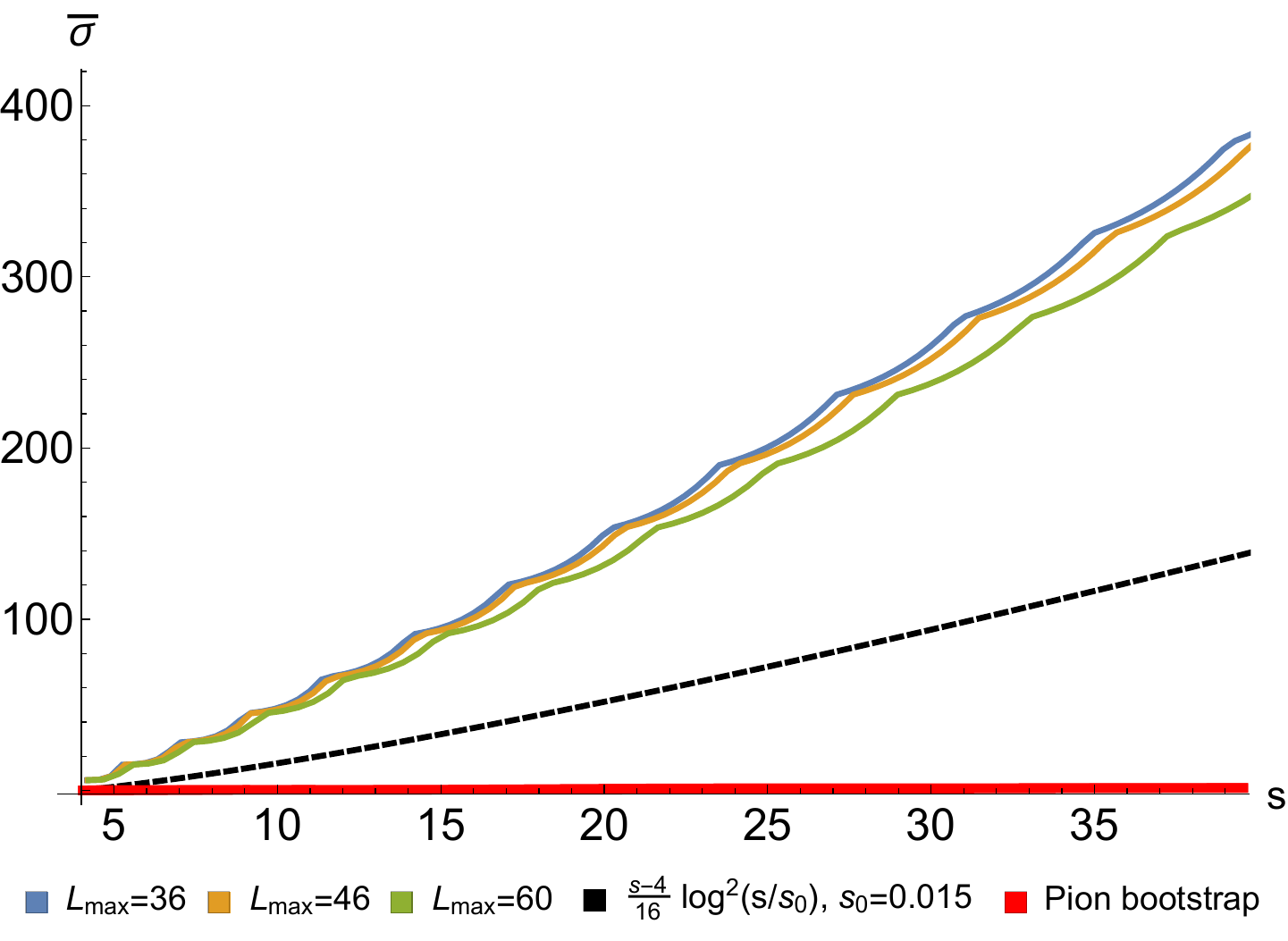}
  \caption{The bound on $\bar{\s}(s)$ using constraints \eqref{sum01} with $L_{max}=36, 46,60$. The black dashed line is the Froissart bound on $\bar{\s}$. The red line is maximum $\bar{\s}$ obtained using the pion bootstrap in \cite{ABPHAS}. }
  \label{fig:Lmax36smax90Nmax480}
\end{figure}
The bound is shown in figure \eqref{fig:Lmax36smax90Nmax480} for various \cite{foot9} $L_{\max}=36,46,60$. A fit with $\bar{\s}_0 \times s \log^2\left(\frac{s}{s_0}\right)$ can be found. The $L_{\max}=36$ gives fit values $\bar{\s}_0=0.29,s_0=0.128$, similarly $L_{\max}=60$ gives fit values $\bar{\s}_0=0.22,s_0=0.066$. The convergence with the spin sum is suggested by the figure but appears to be slow for higher values of $s$.
Also it is clear that the $\bar{\s}$ found using a typical S-matrix living on the boundary of the so-called river arising from the S-matrix bootstrap \cite{ABPHAS} or even the lake boundary in \cite{andrea}, is far below the numerical bound presented.

Note that Froissart bound \cite{foot10} \textit{i.e.,} $\bar{\s}\lesssim \frac{s}{16}~\log^2(s/s_0)$ (which is not valid for lower value of $s$ we are considering) is below the numerical bound. The main utility of our numerical bound is that it is valid for {\it any} $s\geq 4$ unlike the Froissart bound which is valid for $s\gg 4$. It will be fascinating to derive analytic bounds using the present method and see if a stronger than Froissart bound is possible at higher energies. In existing derivations of the Froissart bound, the role of crossing symmetry has not been explored. We derive a simple analytic bound in the supplementary material. \\
{\bf Feynman Block expansion of amplitude: }
We will now address how the structure of the Feynman diagram expansion emerges from our analysis.
Given the partial wave expansion of the s-channel discontinuity, we can write the amplitude as
\begin{small}
\be
\label{eq:Dyson2}
\begin{split}
&\mathcal{M}(s_1, s_2)=\alpha_{0}+\frac{1}{\pi} \int_{\frac{2\m}{3}}^{\infty}\sum_{\ell=0}^{\infty} \frac{d \s}{\s}H\left(\s ;s_1, s_2, s_3\right)\\
&\left( \Phi(\s;\a)\left(2\ell+2\a\right)~\frac{a_\ell(\s)}{\left(\s-\frac{2\m}{3}\right)^\ell}~Q_\ell(\s;s_2^{(+)}(\s,a))\right)\,,
\end{split}
\ee
\end{small}
where $
Q_{\ell}(s_1,s_2)=\left(s_1-\frac{2\m}{3}\right)^\ell C^{(\a)}_{\ell}\left(1+2\frac{s_2+\frac{\m}{3}}{s_1-\frac{2\m}{3}}\right)\,.$ Let us refer to $H Q_\ell$ as the Dyson block--note that this block has non-local negative powers of $x$ which should cancel on using the null constraints. A natural way to proceed would be to use a basis which has these spurious negative powers of $x$ removed at the onset. This leads us to define the Feynman block:
\be
M^F_{\ell}(\s;s_1, s_2)=\sum_{i=1}^{3}M_{\ell}^{(i)}(\s;s_1, s_2)+M_\ell^{(c)}(\sigma;s_1,s_2)\,,\nonumber
\ee
where, s-channel part of the Feynman block is
\be\nonumber
M_{\ell}^{(1)}(\s;s_1, s_2)=Q_{\ell}(s_1, s_2)\left(\frac{1}{{\s-s_1}}-\frac{1}{\s}
\right)\, .
\ee
The $t,u$-channel are given by
$M_{\ell}^{(2)}(\s;s_1, s_2)=M_{ \ell}^{(1)}(\s;s_2, s_3)$,
$M_{\ell}^{(3)}(\s;s_1, s_2)=M_{ \ell}^{(1)}(\s;s_3, s_1)$. $M_{\ell}^{(c)}(\s;s_1, s_2)$ are contact terms involving polynomials of $s_i$'s. To find this write
\be\nonumber
\mathcal{D}_{\ell}(\s)=\frac{1}{\s}Q_{\ell}(\s,s_2^{(+)})H(\s;s_1,s_2,s_3)-\sum_{i=1}^{3}M_{\ell}^{(i)}(\s;s_1, s_2)\,.
\ee
For example
$
\mathcal{D}_{\ell=2}(\s)=\frac{x~ b_{0,2}^{(2)}}{a-\s}+\frac{x~ (b_{0,2}^{(2)}+2 b_{2,0}^{(2)})}{\s}
=\frac{2 x~ b_{2,0}^{(2)}}{\s}-\frac{y~ b_{0,2}^{(2)}}{ (\s)^2}-\left(x~b_{0,2}^{(2)}\right)\sum_{n=2}^{\infty}a^n (\s)^{-n-1}\,,$
where $b_{n,m}^{(\ell)}=\frac{1}{n!m!}\partial_{s_1}^{n}\partial_{s_2}^{m}\left[Q_{\ell}(s_1,s_2)\right]$. We throw away negative powers in $x$ (non-local terms), which gives
\be
M_{\ell=2}^{(c)}(\s;s_1, s_2)=\frac{2 x~ b_{2,0}^{(2)}}{\s}-\frac{y~ b_{0,2}^{(2)}}{ \s^2}\,,
\ee
which is the polynomial form of the contact term. 
This can be repeated for any $\ell$. In general, we can write the Feynman block expansion of an amplitude
\be
\label{eq:Feynman}
\begin{split}
&\mathcal{M}(s_1, s_2)=\alpha_{0}+\frac{1}{\pi} \sum_{\ell=0}^{\infty}\int_{\frac{2\m}{3}}^{\infty} \frac{d \s}{\left(\s-\frac{2\m}{3}\right)^\ell}\\&\times \Big( \Phi(\s;\a)\left(2\ell+2\a\right)~a_\ell(\s)~M^F_\ell(\s;s_1,s_2)\Big)\,,
\end{split}
\ee
{This demonstrates how the structure of the Feynman diagram expansion, involving exchanges and contact diagrams, emerges from the dispersion relation. Eq. \eqref{eq:Dyson2} converges for $-6.71\mu<a<2\mu/3$ \cite{AK}. Thus, expanding \eqref{eq:Dyson2} around $a=0$ gives a convergent representation in $a^n x^m$ powers. The difference between \eqref{eq:Feynman} and  \eqref{eq:Dyson2} are the terms $m<n$ in the latter.  Removing these will leave us with a convergent representation \eqref{eq:Feynman} around $a=0$--see supplementary material for further checks.}
\\
{\bf Discussion: }
The crossing symmetric dispersion relation approach, presented in this paper, promises to open up a new and efficient way to study field theories. We saw how the picture of Feynman diagrams emerges from the crossing symmetric approach; seeing this using the fixed-$t$ dispersion relation is impossible as crossing symmetry is imposed as a constraint. It will be very interesting to connect the ideas and techniques in this paper with the ``EFT-hedron'' picture in \cite{huang}. Another place where we expect these crossing symmetric dispersion relations to play an important role \cite{dualus} is the formulation of the dual S-matrix bootstrap in higher dimensions. So far, an explicit attempt has only been made in 2 dimensions \cite{dual}.

On the CFT side, we have shown in \cite{GSZ} how the manifestly crossing symmetric method extended to CFT Mellin amplitudes leads to the sum rule constraints arising from the two-channel dispersion relation presented in \cite{joaopaper},\cite{joaopaper2}. Furthermore, the CFT generalized null constraints admit a straightforward derivation and are needed to show the equivalence. This suggests that the manifestly crossing symmetric dispersion relation will not only be more systematic but will have more constraints than what is easily derivable in the two-channel symmetric approach.
\section*{Acknowledgments} 
We thank Parijat Dey, Kausik Ghosh, Parthiv Haldar and Apratim Kaviraj for useful discussions. Special thanks to Andrew Tolley for comments on the draft and to Yu-tin Huang for comments on the draft and sharing \cite{huang}. We especially thank Rajesh Gopakumar for numerous discussions and encouragement for this work, comments on the draft, as well as for ongoing collaboration in \cite{GSZ}.


\onecolumngrid
{\begin{center}\bf \Large{Supplementary material}\end{center} }

\section{Crossing symmetric dispersion relation}\label{ap:crosscutandall}
This section reviews the derivation of a crossing symmetric dispersion relation based on \cite{AK}. In what follows we use $\overline{\mathcal{M}} (z,a)$ to distinguish from $\mathcal{M}(s_1,s_2)$.
\\

\textbf{Image of the physical cuts:}
\label{ap:cutVaderivation}
The image of the three physical cuts in the $z$-plane ,  $V(a)=V_{1}(a) \cup V_{2}(a) \cup V_{3}(a)$ can be derived from eq (1) in the main text. For example, we want $s_1\geq \frac{2\m}{3}$ (or $s\geq \m$), which is only possible if $\text{Im}\left(s_1 \right)=0$. Now  $\text{Im}\left(s_1 \right)=0$ if $a=0$ or $|z|=1$ or $\arg(z)=\pi$. Let us take the case of $|z|=1$, then conditions $s_1\geq \frac{2\m}{3}$ and $-\frac{\m}{3}\leq a<\frac{2\m}{3}$ constrain the value of $\arg(z)$. For $-\frac{2\m}{9}< a<0$ $$\frac{2\pi}{3}\leq \arg(z)\leq \phi_0(a)\,,$$
where $\phi_0(a)$ is given below and  for $0<a<\frac{2\m}{3}$ $$\phi_0(a)\leq \arg(z)\leq \frac{2\pi}{3}\,.$$ 
Another case of $\arg(z)=\pi$ can be easily worked out similarly, which leads to \cite[eq (3.11)]{AK}
\be\label{eq:cutVa}
\begin{split}
&V_1(a)=\left\{\begin{array}{l}
\left\{z; | z|=1, \frac{2}{3} \pi \leqslant | \arg z \mid \leqslant \phi_{0}(a)\right\} \quad \text { if } -\frac{2\m}{9}<a<0\quad\quad\quad(\mathrm{I})\,,\\
\left\{z; | z\left|=1, \phi_{0}(a) \leqslant\right| \arg z \mid \leqslant \frac{2}{3} \pi\right\} \quad \text {~ if } 0<a<\frac{2\m}{3}\quad\quad \quad\quad(\mathrm{II})\,,\\
\left\{z; | z\left|=1, \frac{2}{3} \pi \leqslant\right| \arg z \mid \leqslant \pi\right\} \cup \left\{z\left|\rho_{-}(a) \leqslant\right| z \mid \leqslant \rho_{+}(a), \arg z=\pi\right\} ~ \text{if } a \leqslant -\frac{2\m}{9} ~(\mathrm{III})\,,
\end{array}\right. \\
&\begin{array}{l}
V_{2}(a)=e^{2 i \pi / 3} V_{1}(a), ~V_{3}(a)=e^{4 i \pi / 3} V_{1}(a)\,,
\end{array}
\end{split}
\ee
with
\be
\begin{aligned}
&\phi_{0}(a)=\tan ^{-1}\left\{\frac{\left[(\frac{2\m}{3}-a)\left(a+\frac{2\m}{9}\right)\right]^{1 / 2}}{a-\frac{2\m}{9}}\right\}, \quad 0<\phi_{0} \leqslant \pi\\
&\rho_{\pm}(a)=\frac{9}{4\m}\left\{\left(\frac{2\m}{9}-a\right) \pm\left[(\frac{2\m}{3}-a)\left(-a-\frac{2\m}{9}\right)\right]^{1 / 2}\right\}\,.
\end{aligned}
\ee

\begin{figure}[hbt!]
  \begin{subfigure}{5cm}
    \centering\includegraphics[width=5cm]{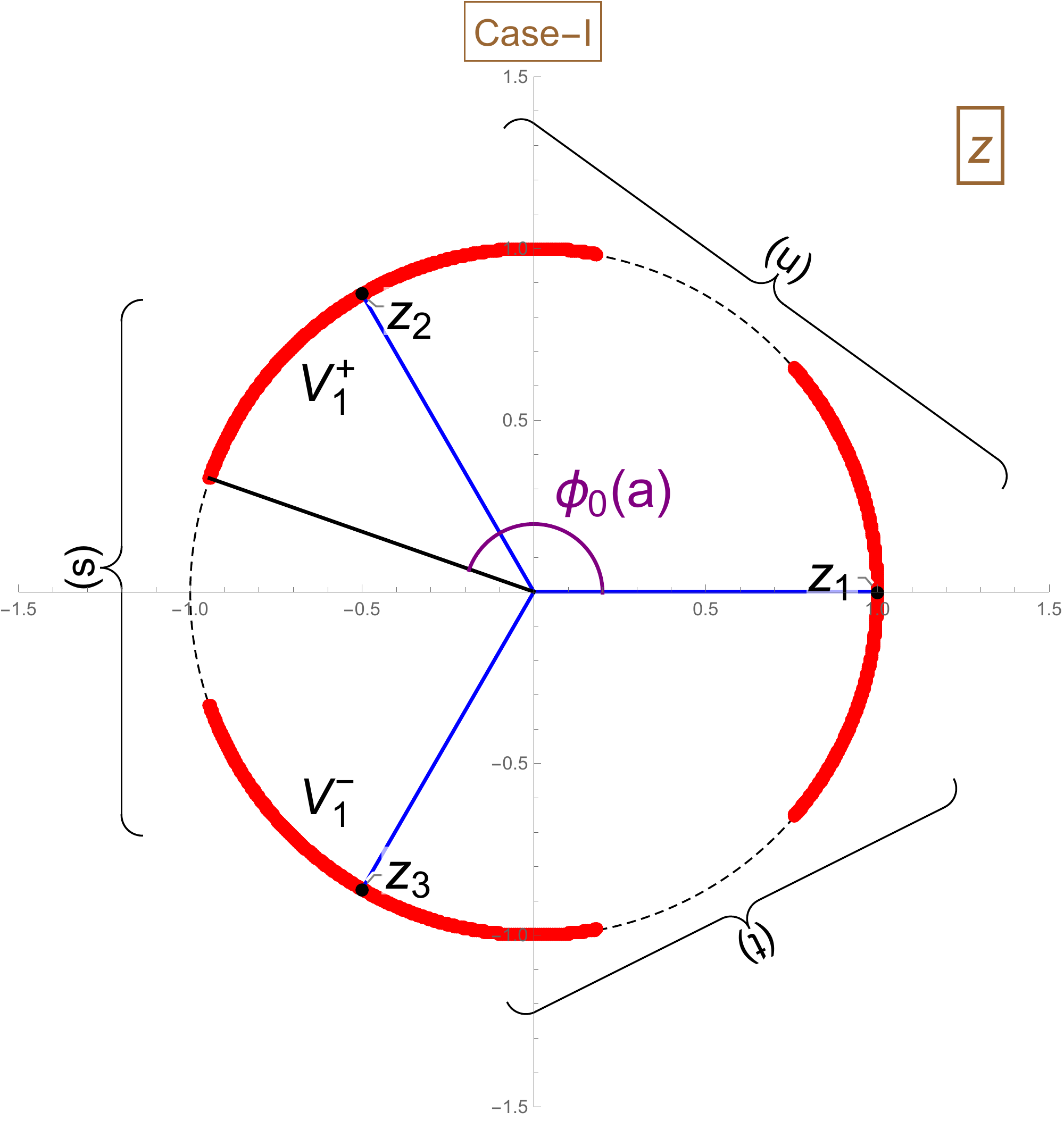}
    \caption{Case-I}
  \end{subfigure}
  \begin{subfigure}{5cm}
    \centering\includegraphics[width=5cm]{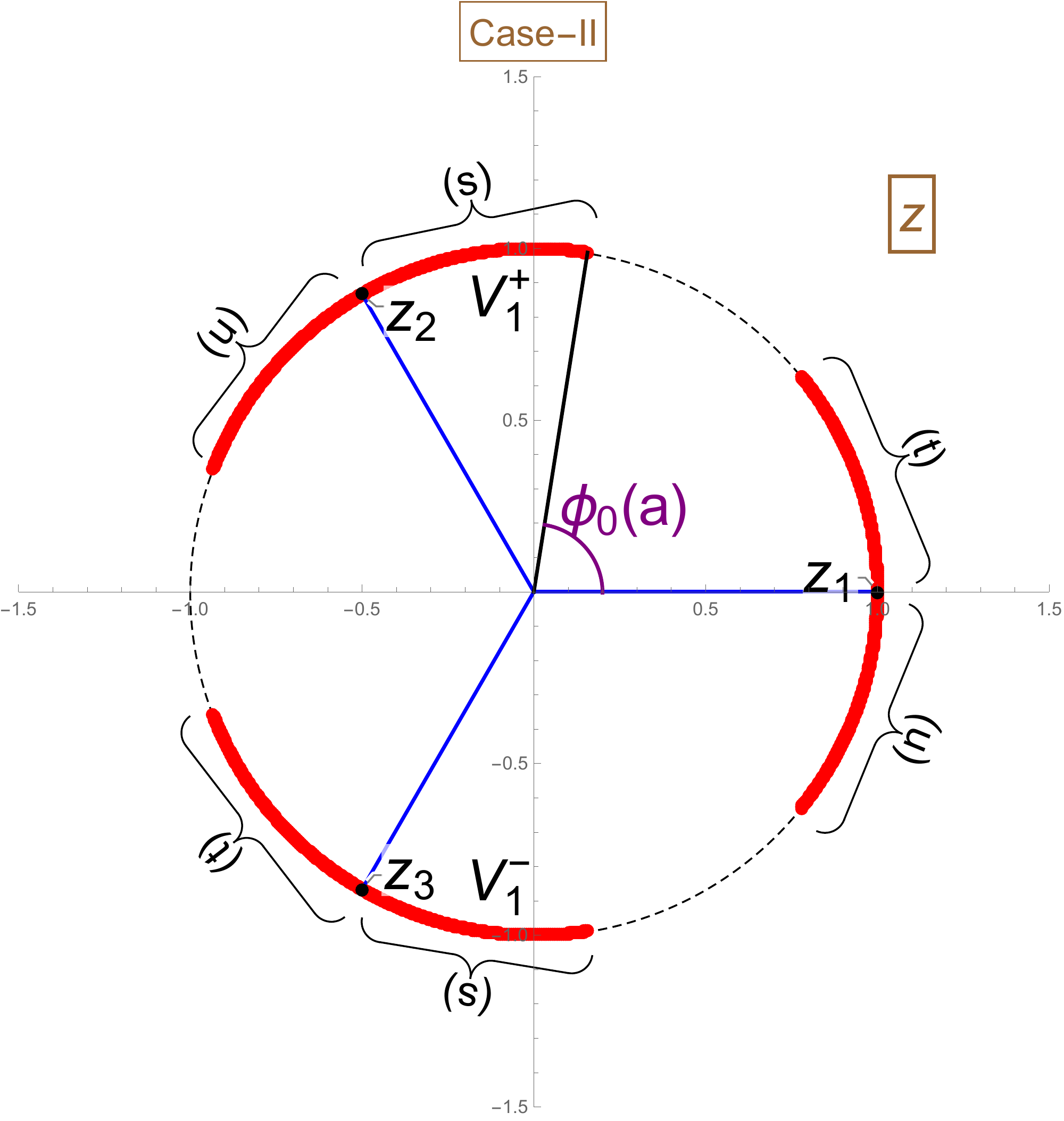}
    \caption{Case-II}
  \end{subfigure}
  \begin{subfigure}{5cm}
    \centering\includegraphics[width=5cm]{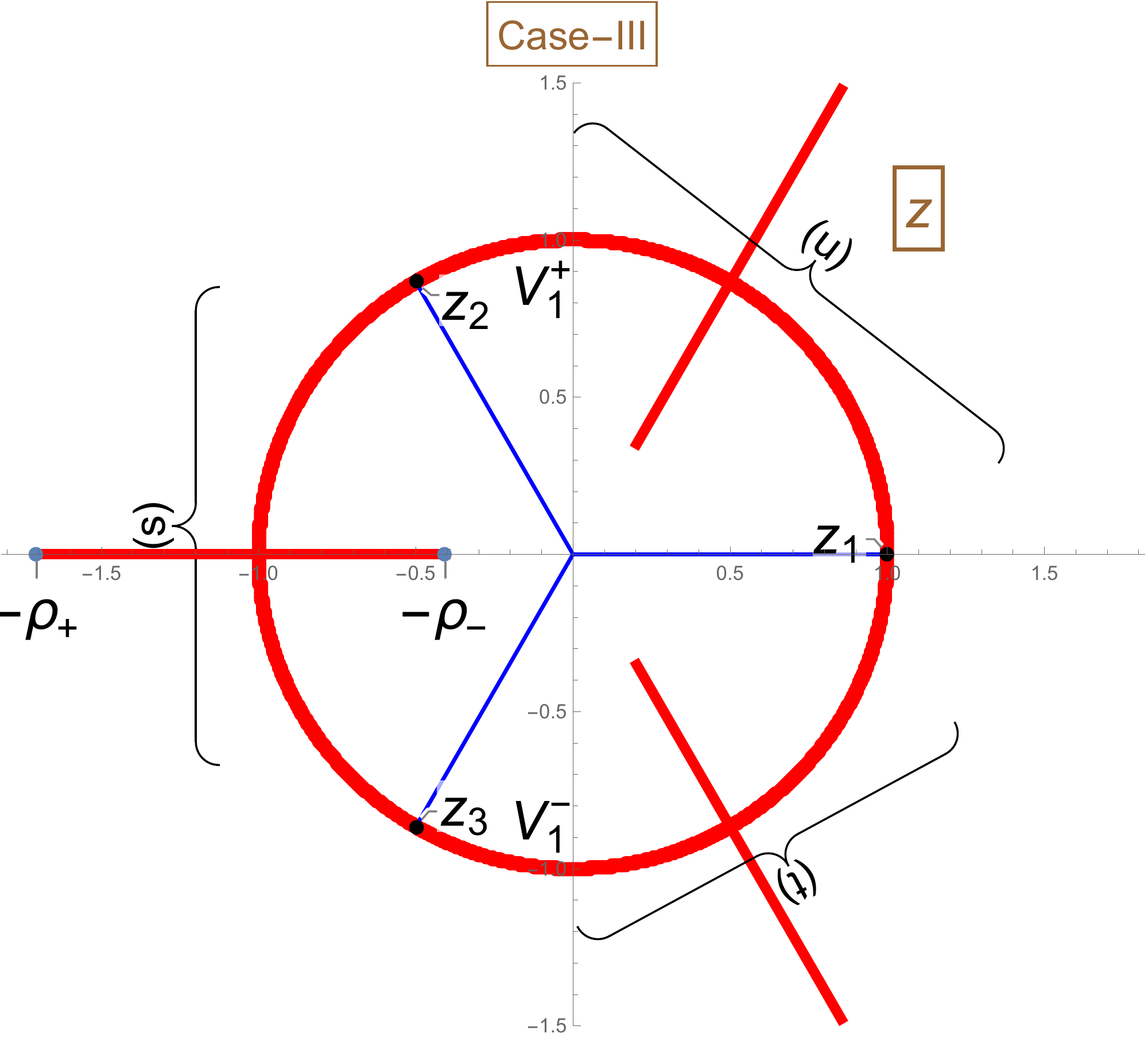}
    \caption{Case-III}
  \end{subfigure}
  \caption{All three cases of $V(a)$. In the above plot we have assumed $\m=4$.}
  \label{fig:Case123}
\end{figure}

\textbf{Parametric Dispersion Relation:}
\label{ap:dispersionderivationgen}
We denote the $k$-channel discontinuity of $\mathcal{M}(s_1,s_2)$ as $\mathcal{A}_{k}(s_1,s_2)$. The usual definition, for example, in the $s$-channel is
\be\label{eq:discontinuity1}
\mathcal{A}_{1}(s_1, s_2) \equiv \lim _{\epsilon \to 0} \frac{1}{2 i}[\mathcal{M}(s_1+i \epsilon, s_2)-\mathcal{M}(s_1-i \epsilon, s_2)], \quad s \geqslant 2\m/3\,.
\ee
In a similar fashion, the discontinuity across $V(a)$ is defined by
\be\label{eq:discontinuityz}
\begin{split}
\overline{\mathcal{A}}(z, a) \equiv\left\{\begin{array}{ll}
\lim _{\epsilon \to 0} \frac{1}{2 i}[\overline{\mathcal{M}}((1+\epsilon) z, a)-\overline{\mathcal{M}}((1-\epsilon) z, a)] & \text { for }|z|=1 \\
\lim _{\epsilon \to 0} \frac{1}{2 i}\left[\overline{\mathcal{M}}\left(z e^{-i \epsilon}, a\right)-\overline{\mathcal{M}}\left(z e^{i \epsilon}, a\right)\right] & \text { for arg } z=\pi\left(\bmod \frac{2}{3} \pi\right)
\end{array}\right.
\end{split}
\ee
The condition $\mathcal{M}(s,t)=\mathcal{M}^*(s^*,t^*)$ translates in to 
$
\overline{\mathcal{M}}(z,a)=\overline{\mathcal{M}}^*(1/z^*,a)\,,
$
which directly follows from eq (1) of main text. This relation implies
\be
\begin{split}
&\overline{\mathcal{A}}(z, a)=\overline{\mathcal{A}}^*(z, a) \quad \text { for }|z|=1\\
&\overline{\mathcal{A}}(z, a)=-\overline{\mathcal{A}}^*\left(\frac{1}{z^{*}}, a\right) \quad \text { for } \arg z=\pi\left(\bmod \frac{2}{3} \pi\right)\,.
\end{split}
\ee
In order to relate  $\overline{\mathcal{A}}(z, a)$ to $\mathcal{A}_k(s,t)$, we first note that $V(a)$ can be split into two subsets depending on the sign of $\text{Im}(s_k)$, namely
\be 
V_{k}(a)=V_{k}^{+}(a) \cup V_{k}^{-}(a), \quad \text { with } \quad \operatorname{Im}\left(z / z_{k}\right) \quad \text { or } \quad(|z|-1)\left\{\begin{array}{l}
\geqslant 0 \text { in } V_{k}^{+}(a) \\
\leqslant 0 \text { in } V_{k}^{-}(a)
\end{array}\right.
\ee
If $z$ is in $V_{k}^{+}(a)$ then the sign of $\text{Im}(s_k)$ is positive and vice-versa, which implies
\begin{small}
\be
\begin{aligned}
&\overline{\mathcal{A}}(z, a)=\left\{\begin{array}{ll}
\mathcal{A}_{k}(s_1, s_2), & z \in V_{k}^{+}(a) \\
-\mathcal{A}_{k}(s_1, s_2), & z \in V_{k}^{-}(a)
\end{array} \quad a<0\right. \text { (cases I and III) }\\
&\overline{\mathcal{A}}(z, a)=\left\{\begin{array}{ll}
-\mathcal{A}_{k}(s_1, s_2), & z \in V_{k}^{+}(a) \\
\mathcal{A}_{k}(s_1, s_2), & z \in V_{k}^{-}(a)
\end{array} \quad \begin{array}{l}
a>0 \quad \text { (case II) }
\end{array}\right.
\end{aligned}\,.
\ee
\end{small}
Following the standard practice of allowing for at most two subtractions \cite{Martin},
\be
\mathcal{M}(s_1, s_2)=o\left(s_1^{2}\right) \text { for }|s_1|\to\infty, \quad s_2 =\text { fixed }, \quad(s_1, s_2) \in \mathcal{D}
\ee
Here $\mathcal{D}$ denotes the Martin domains \cite{AK, Martin}.
Similarly for fixed $s_1,s_3$. We want to study a similar behaviour of $\overline{\mathcal{M}}(z,a)$ around $z\to z_k$. We notice that 
$
s_{k}\to a, \quad s_{j} \simeq \operatorname{constant}\times \frac{a}{z-z_{k}} \quad(j \neq k)\,.
$ Therefore, we have
\be\label{eq:Mbarfall}
\overline{\mathcal{M}}(z, a)=o\left(\frac{1}{\left(z-z_{k}\right)^{2}}\right), \quad \text { as } \quad z\to z_{k}, \quad a \text { fixed }
\ee
We Taylor expand the amplitude (convergent for $|z|<\r_-(a)$ or $|z|<1$)
\begin{small}
\be 
\overline{\mathcal{M}}(z, a)=\sum_{n=0}^{\infty} f_{n}(a) z^{n}\,.
\ee
\end{small}
Note that 
$
\overline{\mathcal{M}}(0, a)=\mathcal{M}\left(s_1=0, s_2=0\right)=f_0=\overline{\mathcal{M}}(\infty, a)=\overline{\mathcal{M}} ^*(0, a)\,,
$
therefore $f_0$ is real and independent of $a$. We can write
\begin{small}
\be
\begin{split}
&\frac{1}{2 \pi i} \oint_{\mathcal{I}} d z^{\prime} \frac{z^{\prime 3}-1}{z^{\prime 3}\left(z^{\prime}-z\right)} \overline{\mathcal{M}}\left(z^{\prime}, a\right)=\frac{z^{3}-1}{z^{3}} \overline{\mathcal{M}}(z, a)+\frac{f_{0}}{z^{3}}+\frac{f_{1}(a)}{z^{2}}+\frac{f_{2}(a)}{z} \,,\\
&\frac{1}{2 \pi i} \oint_{\mathcal{E}} d z^{\prime} \frac{z^{\prime 3}-1}{z^{\prime 3}\left(z^{\prime}-z\right)} \overline{\mathcal{M}}\left(z^{\prime}, a\right)=\overline{\mathcal{M}}(\infty, a)=f_{0}\,,
\end{split}
\ee
\end{small}
where $\mathcal{I}, \mathcal{E}$ are the interior and exterior of $V(a)$ with $z \notin V(a), |z|<1$. In case of $\mathcal{I}$, the terms involving $f_0,f_1,f_2$ follows from the residue at $z=0$, and the term $\frac{z^{3}-1}{z^{3}} \overline{\mathcal{M}}(z, a)$ comes from residue at $z=z'$. In case of $\mathcal{E}$, we can take the contour upto infinity. We can subtract these two equations on $\mathcal{I}, \mathcal{E}$. Now as $\mathcal{I}, \mathcal{E}$  approaches to $V(a)$, we get the dispersion relation across $V(a)$:
\be
\label{eq:para_dis}
\overline{\mathcal{M}}(z, a)=f_{0}+f_{1}(a) \frac{z}{1-z^{3}}+f_{2}(a) \frac{z^{2}}{1-z^{3}}+\frac{z^{3}}{\left(1-z^{3}\right)} \frac{1}{\pi} \int_{V(a)} d z^{\prime} \frac{z^{\prime 3}-1}{z^{\prime 3}\left(z^{\prime}-z\right)} \overline{\mathcal{A}}\left(z^{\prime}, a\right)\,
\ee
where $\overline{\mathcal{A}}\left(z^{\prime}, a\right)$ is the discontinuity of the amplitude across $V(a)$, defined in eq \eqref{eq:discontinuityz}.

Now we consider the completely crossing symmetric case (for example $\pi^0\pi^0$ scattering). In such a case $\overline{\mathcal{M}}(z,a)$ should be a function of $z^3$.
\be\label{eq:M0_exp}
\overline{\mathcal{M}}_0(z, a)=\sum_{n=0}^{\infty} \alpha_{n}(a) z^{3 n}\,.
\ee
Therefore the dispersion relation on $V(a)$, \textit{i.e.} eq \eqref{eq:para_dis} becomes 
\be\label{eq:M0_disper}
\overline{\mathcal{M}}_0(z, a)=\alpha_{0}+\frac{z^{3}}{\left(1-z^{3}\right) \pi} \int_{V(a)} d z^{\prime} \frac{z^{\prime 3}-1}{z^{\prime 3}\left(z^{\prime}-z\right)} \overline{\mathcal{A}}\left(z^{\prime}, a\right)\,.
\ee
In the above equation, we are only interested in powers like $z^{3n},~n\in\mathbb{N}$ . While other powers like $z^m, m\notin 3\mathbb{N}$ should disappear from the RHS of \eqref{eq:M0_disper}. We achieve this as follows. Comparing \eqref{eq:M0_disper} with \eqref{eq:M0_exp}, we find
\be\label{eq:alpha}
\alpha_{n}(a)=\frac{1}{\pi} \int_{V(a)} d z^{\prime} \overline{\mathcal{A}}\left(z^{\prime}, a\right) \frac{1-z^{\prime-3 n}}{z^{\prime}}, \quad n \geqslant 1 \,.
\ee
We put the above \eqref{eq:alpha} formula in \eqref{eq:M0_exp}, and doing the sum over $n$,  we get
\be
\label{eq:M0_disper2}
\overline{\mathcal{M}}_0(z, a)=\alpha_{0}+\frac{z^{3}}{\left(1-z^{3}\right) \pi} \int_{V(a)} d z^{\prime} \frac{z^{\prime 3}-1}{z^{\prime }\left(z^{\prime 3}-z^3\right)} \overline{\mathcal{A}}\left(z^{\prime}, a\right)\,.
\ee
While writing the above equation in $s_1,~s_2$ variable, we follow the method as follows (for $-2\m/9<a<0$). We solve for $z^{\prime}$ in terms of $s_1'$. Notice that $z^{\prime}$ has two solutions in terms of $s_1'$. If one solution is $z^{\prime}(s_1')$ then other solution is $z^{*\prime}(s_1')$, also $|z^{\prime}(s_1')|=1$. First two terms are given by
\be
\frac{z^{3}}{\left(1-z^{3}\right) }  \frac{d z^{\prime}(s_1')}{ds_1'} \frac{z^{\prime 3}(s_1')-1}{z^{\prime}(s_1')\left(z^{\prime 3}(s_1')-z^3\right)} \overline{\mathcal{A}}\left(z^{\prime}(s_1'), a\right)+\frac{z^{3}}{\left(1-z^{3}\right) }  \frac{d z^{*\prime}(s_1')}{ds_1'} \frac{z^{*\prime 3}(s_1')-1}{z^{*\prime}(s_1')\left(z^{*\prime 3}(s_1')-z^3\right)} \overline{\mathcal{A}}\left(z^{*\prime}(s_1'), a\right)\,.
\ee
Now we note that $\overline{\mathcal{A}}\left(z^{\prime}(s_1'), a\right)=\overline{\mathcal{A}}\left(z^{*\prime}(s_1'), a\right)=\mathcal{A}_1\left(s_1',s_2^{+}(s_1',a)\right)$. One can easily see that $\overline{\mathcal{M}}_0\left(z, a\right)$ is function of combinations of $\frac{z^3}{(z^3-1)^2}$. This property follows from complete crossing symmetry, then $\overline{\mathcal{A}}\left(z, a\right)$ should also have this property. We also note $\frac{z^3}{(z^3-1)^2}$ is invariant under $z\to \frac{1}{z}$ and  $z*=\frac{1}{z}$ for $|z|=1$. Now we note that $z'=\exp (i \phi )$ with $\phi= \cos ^{-1}\left(\frac{3 a-s_1'}{2 s_1'}\right)$.    We finally have (eq (2) in main text).
\be
\begin{split}
\mathcal{M}_{0}(s_1, s_2)=\alpha_{0}+\frac{1}{\pi} \int_{\frac{2\m}{3} }^{\infty} \frac{d s_1^{\prime}}{s_1^{\prime}} \mathcal{A}\left(s_1^{\prime} ; s_2^{(+)}\left(s_1^{\prime} ,a\right)\right) H\left(s_1^{\prime} ;s_1, s_2, s_3\right)\,,
\end{split}
\ee
where $s_2^{(+)}\left(s_1^{\prime} ,a\right)$ and $H\left(s_1^{\prime} ;s_1, s_2, s_3\right)$ are given in the main text.
\\
\textbf{Comparison with fixed-$s_2$ dispersion relations:}
Given s-channel discontinuity the closed superstring tree amplitude, one can calculate  of the amplitude via crossing symmetric dispersion relations as well as via fixed-$s_2$ dispersion relations. We will consider
\be\nonumber
-s_1 s_2 (s_1+s_2)\mathcal{M}(s_1,s_2)=\frac{\Gamma \left(1-s_1\right) \Gamma \left(1-s_2\right) \Gamma \left(s_1+s_2+1\right)}{\Gamma \left(s_1+1\right) \Gamma \left(-s_1-s_2+1\right) \Gamma \left(s_2+1\right)}\,.
\ee
From the crossing symmetric dispersion relation, we get 
\begin{small}
\be\label{eq:cross_string}
\begin{split}
&-s_1 s_2 (s_1+s_2)\mathcal{M}(s_1,s_2)^{(crossing)}=1+\sum_{k=0}^{\infty}\Bigg[\frac{(-1)^k}{k!(k+1)!}\left(\frac{1}{k-s_1+1}+\frac{1}{k-s_2+1}+\frac{1}{k-s_3+1}-\frac{3}{k+1}\right)\\
&\times\frac{\Gamma \left(\frac{1}{2} \left(-k~\sqrt{\frac{4 a}{-a+k+1}+1} +k-\sqrt{\frac{4 a}{-a+k+1}+1}+3\right)\right) \Gamma \left(\frac{1}{2} \left(k~\sqrt{\frac{4 a}{-a+k+1}+1} +k+\sqrt{\frac{4 a}{-a+k+1}+1}+3\right)\right)}{\Gamma \left(\frac{1}{2} \left(k \left(\sqrt{\frac{4 a}{-a+k+1}+1}-1\right)+\sqrt{\frac{4 a}{-a+k+1}+1}+1\right)\right) \Gamma \left(\frac{1}{2} \left(-\sqrt{\frac{4 a}{-a+k+1}+1}-k \left(\sqrt{\frac{4 a}{-a+k+1}+1}+1\right)+1\right)\right)}\Bigg]\,,
\end{split}
\ee
\end{small}
and from 2-channel dispersion relation, we get
\begin{small}
\be\label{eq:2ch_string}
\begin{split}
&-s_1 s_2 (s_1+s_2)\mathcal{M}(s_1,s_2)^{(2-channel)}=1+\sum_{k=0}^{\infty}\Bigg[\frac{(-1)^{k+1} s_1 \left(s_1+s_2\right) \left(2 k+s_2+2\right) \Gamma \left(1-s_2\right) \Gamma \left(k+s_2+1\right)}{\left(k-s_1+1\right) \left(k+s_1+s_2+1\right) \Gamma (k+2)^2 \Gamma \left(s_2+1\right) \Gamma \left(-k-s_2\right)}\Bigg]\,.
\end{split}
\ee
\end{small}\\
Various comparisons are shown in figure \eqref{fig:comparison}. In figure \ref{fig:comparison} (a), we have shown that the region of $s_1, s_2$, where the dispersion relations are within $10\%$ accuracy compared to the actual amplitude. The overlaps of the regions where both dispersion relations give $10\%$ accuracy are also shown \ref{fig:comparison} (a). In figure  \ref{fig:comparison} (b), we have shown the amplitude for fixed $s_2=\frac{1}{13}$ as a function of $s_1$. {Notice that the crossing symmetric expression in \eqref{eq:cross_string}, contains non-physical singularities as a function of $a$. These poles are responsible for the negative powers of $x,y$. Cancellation of such terms leads to the ``null constraints". As we sum over $k$ these singularities add up to zero. 
 }
\begin{figure}[ht]
  \begin{subfigure}{8.5cm}
  \centering\includegraphics[width=8.5cm]{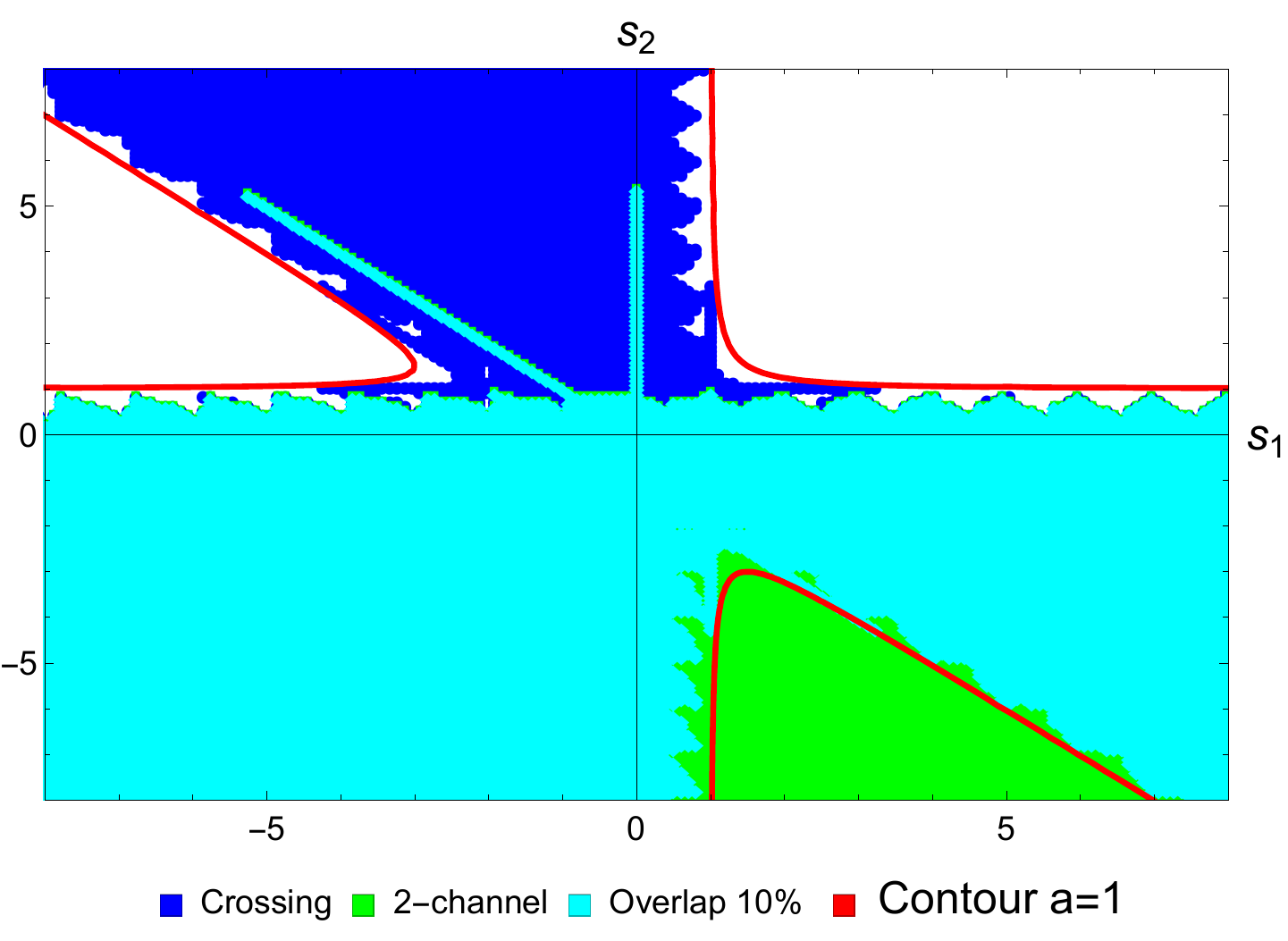}
    \caption{Region of $s_1, s_2$ where the dispersion relations gives $10\%$ accuracy, in crossing symmetric dispersion (Blue) as well as fixed-$s_2$ dispersion relation (Green). Overlap region of $10\%$ accuracy (Cyan).}
  \end{subfigure}
  \begin{subfigure}{8.5cm}
   \centering\includegraphics[width=8.5cm]{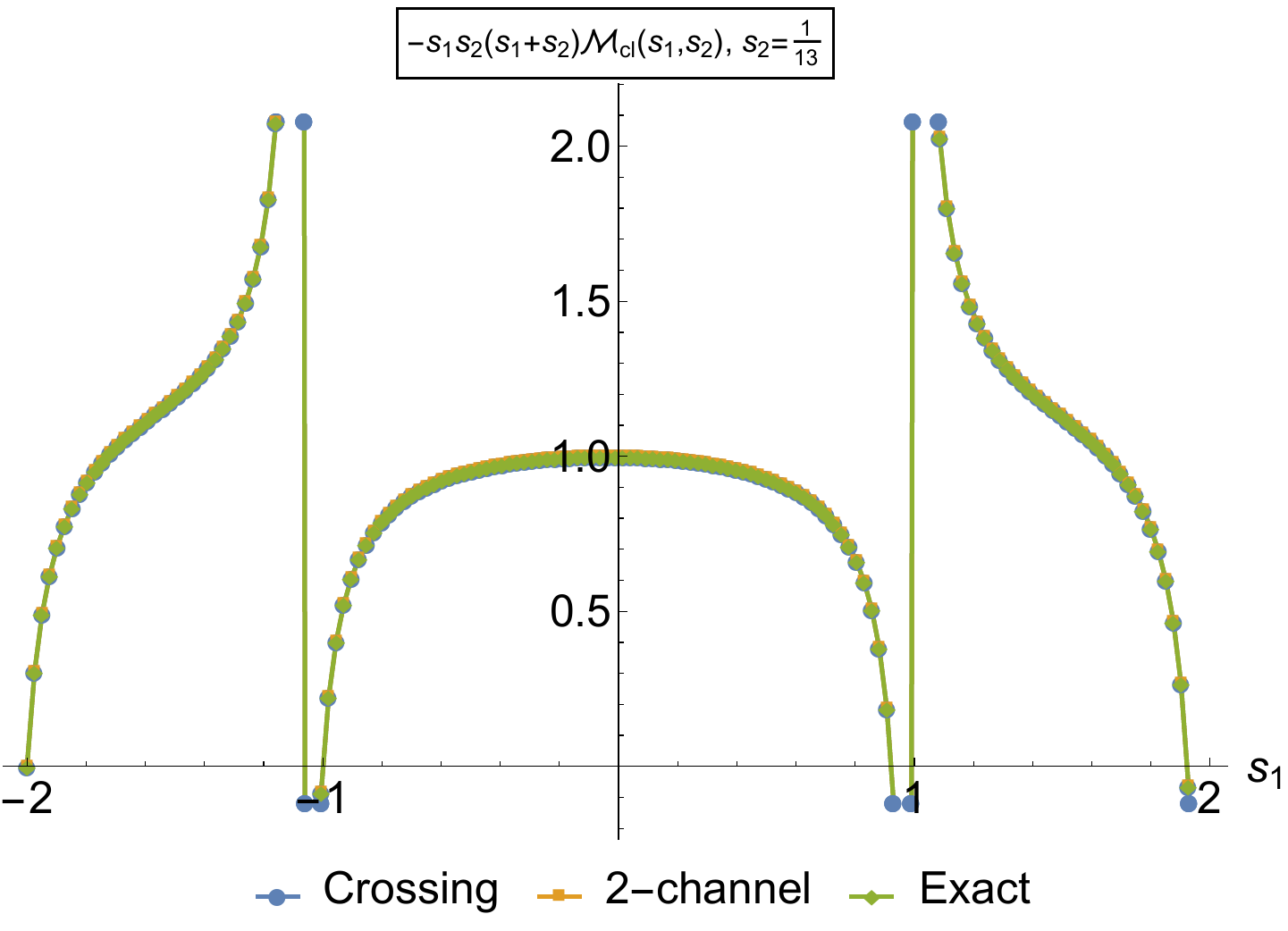}
   \caption{Comparison of amplitude for $s_2=\frac{1}{13}$ calculated using crossing symmetric dispersion relations and fixed-$s_2$ dispersion relations with the exact amplitude.}
  \end{subfigure}
  \caption{Comparison between crossing symmetric dispersion relation and fixed-t dispersion relation. We have truncated the $k-$ sum upto $k_{\max}=100$ }
  \label{fig:comparison}
\end{figure}
\\
\textbf{\emph{Some comments:}}
We note that we have written a dispersion relation for $\frac{z^{3}-1}{z^{3}} \overline{M}(z,a)$ to cure the $o\left(\frac{1}{(z-z_k)^2}\right)$ behaviour of the amplitude around $z\to z_k$ presented in \eqref{eq:Mbarfall}. If an amplitude does not have this behaviour, say it has a behaviour $o\left(\frac{1}{(z-z_k)}\right)$, then it is possible to write a dispersion relation for $\overline{M}(z,a)$ without the $\frac{z^{3}-1}{z^{3}}$ factor. More interestingly, we find that if the amplitude has a behaviour like $o\left(\frac{1}{(z-z_k)^0}\right)$ then we can write down 
\begin{small}
\be\nonumber
\begin{split}
\frac{1}{2 \pi i} \oint_{\mathcal{I}} d z^{\prime} \frac{3 \left(\left(z'\right)^2-1\right) \left(z^3+z' \left(z'+z^6 z'+z^3 \left(z'+2\right) \left(\left(z'\right)^2+1\right)\right)\right)}{\left(\left(z'\right)^2+z'+1\right) \left(z^3-\left(z^6+1\right) \left(z'\right)^3+z^3 \left(z'\right)^6\right)}\overline{\mathcal{M}}\left(z^{\prime}, a\right)=& \sum_{k=0}^{3}\left(\overline{\mathcal{M}}(z_k~z, a)+\overline{\mathcal{M}}(\frac{z_k}{z}, a)\right)\\
&-3\left(\overline{\mathcal{M}}(z_1, a)+\overline{\mathcal{M}}(z_2, a)\right)\,,\\
\frac{1}{2 \pi i} \oint_{\mathcal{E}} d z^{\prime} \frac{3 \left(\left(z'\right)^2-1\right) \left(z^3+z' \left(z'+z^6 z'+z^3 \left(z'+2\right) \left(\left(z'\right)^2+1\right)\right)\right)}{\left(\left(z'\right)^2+z'+1\right) \left(z^3-\left(z^6+1\right) \left(z'\right)^3+z^3 \left(z'\right)^6\right)} \overline{\mathcal{M}}\left(z^{\prime}, a\right)=&0\,,
\end{split}
\ee
\end{small}\\
which gives a dispersion relation of the kind (we have used the fact that $\overline{\mathcal{M}}(z_k~z, a)=\overline{\mathcal{M}}(z, a)$ since it is a function of $z^3$ and $\overline{\mathcal{M}}(z_k, a)=0$)
\be
\begin{split}
\mathcal{M}_{0}(s_1, s_2)=\frac{1}{\pi} \int_{\frac{2\m}{3} }^{\infty} d s_1^{\prime} \mathcal{A}\left(s_1^{\prime} ; s_2^{(+)}\left(s_1^{\prime} ,a\right)\right) H_2\left(s_1^{\prime} ;s_1, s_2, s_3\right)
\end{split}
\ee 
where  
\begin{small}
\be
H_2\left(s_1^{\prime} ;s_1, s_2, s_3\right)=\left[\frac{1}{\left(s_1^{\prime}-s_1\right)}+\frac{1}{\left(s_1^{\prime}-s_2\right)}+\frac{1}{\left(s_1^{\prime}-s_3\right)}\right]\,,
\ee
\end{small}
which is an ``unsubtracted'' dispersion relation.

\section{Some algebraic details}

\textbf{Derivation of eq (7) in main text:}\label{ap:proofofBell2}
We note that
\begin{small}
\be
\begin{split}
&\mathcal{B}^{(\ell)}_{n,1}\left(\d+\frac{2\m}{3}\right)+\chi^{(0,1)}(\m,\d_0)\mathcal{B}^{(\ell)}_{n,0}\left(\d+\frac{2\m}{3}\right)\\
&=\frac{\left(\delta -\delta _0\right) 9^{n+1} (2 n+1) (3 \delta +2 \mu )^{-2 n-1}}{\pi  \left(3 \delta _0+2 \mu \right)}C_{\ell }^{(\alpha)}\left(\frac{2 \mu }{3 \delta }+1\right)+\frac{8 \alpha  \left(\delta +\frac{2 \mu }{3}\right)^{-2 n}}{\pi  \delta }C_{\ell-1 }^{(\alpha+1)}\left(\frac{2 \mu }{3 \delta }+1\right)\geq 0
\end{split}
\ee
\end{small}
for $\d\geq \d_0$. Similarly
\begin{small}
\be
\begin{split}
&\mathcal{B}^{(\ell)}_{n,2}\left(\d+\frac{2\m}{3}\right)+\chi^{(1,2)}(\m,\d_0)\mathcal{B}^{(\ell)}_{n,1}\left(\d+\frac{2\m}{3}\right)+\chi^{(0,2)}(\m,\d_0)\mathcal{B}^{(\ell)}_{n,0}\left(\d+\frac{2\m}{3}\right)\\
&=\frac{\left(\delta -\delta _0\right) 3^{2 n+3} (3 \delta +2 \mu )^{-2 (n+1)}}{2 \pi  \left(3 \delta _0+2 \mu \right){}^2}\Bigg(6 \delta _0+3 \delta  (4 n+3)+2 \mu  (4 n+5)+6n^2(\d-\d_0) \Bigg)C_{\ell }^{(\alpha)}\left(\frac{2 \mu }{3 \delta }+1\right)\\
&+\frac{4 \alpha  \left(\delta -\delta _0\right) 9^{n+1} (2 n+1) (3 \delta +2 \mu )^{-2 n-1}}{\pi  \delta  \left(3 \delta _0+2 \mu \right)} C_{\ell-1 }^{(\alpha+1)}\left(\frac{2 \mu }{3 \delta }+1\right)+\frac{16 \alpha  (\alpha +1) 9^n (3 \delta +2 \mu )^{-2 n}}{\pi  \delta ^2}C_{\ell-2 }^{(\alpha+2)}\left(\frac{2 \mu }{3 \delta }+1\right) \geq 0
\end{split}
\ee
\end{small}
for $\d\geq\d_0$. Similar strategy works up to $m=4$. For $m=5$ onwards we have verified these inequalities numerically.

\section{An analytic bound on total cross-section for any $s>4$}
In this section we use the $m=2,n=1$ null constraint to derive an analytic bound on the averaged cross section. A useful fact (recall we are using $s=s_1+\frac{4}{3}$) that can be established by observation is that ${\mathcal B}_{2,1}^{(\ell_1)}(s)<{\mathcal B}_{2,1}^{(\ell_2)}(s)$ for $4\leq\ell_1<\ell_2$ and $s\geq 4$. In the $m=2,n=1$ case, ${\mathcal B}_{2,1}^{(\ell)}(s)>0$ for $\ell\geq 4$ while ${\mathcal B}_{2,1}^{(2)}(s)<0$ inside the integrand. Rewriting the null constraint as 
\be
\int_4^\infty \frac{ds}{s-\frac{4}{3}}\sqrt{\frac{s}{s-4}} \sum_{\ell=4}^{\infty}(2\ell+1)a_\ell(s) {\mathcal B}_{2,1}^{(\ell)}(s)=5\int_4^\infty \frac{ds}{s-\frac{4}{3}}\sqrt{\frac{s}{s-4}} a_2(s){\bigg |} {\mathcal B}_{2,1}^{(2)}(s){\bigg |}\,.
\ee
Now each term in the summand on the LHS is positive. Truncating the integral to $s=s_{max}$ and substituting for ${\mathcal B}_{2,1}^{(2)}(s)$ we find the inequality:
\be \label{ineqq}
\int_4^{s_{max}} \frac{ds}{s-\frac{4}{3}}\sqrt{\frac{s}{s-4}} \sum_{\ell=4}^{\infty}(2\ell+1)a_\ell(s) {\mathcal B}_{2,1}^{(\ell)}(s)<\frac{30}{\pi}\int_4^\infty ds \sqrt{\frac{s}{s-4}} \frac{a_2(s)}{(s-4)^2(s-\frac{4}{3})^3}\equiv \lambda\,.
\ee
We have assumed that $\lambda$ is finite--the seemingly problematic lower limit of the integral can be shown to be finite using the results in \cite{sasha2}.
Now on the LHS, using the above mentioned monotonicity property for ${\mathcal B}_{2,1}^{(\ell)}(s)$ we can pull out ${\mathcal B}_{2,1}^{(4)}(s_{max})$ from the integral leading to the following bound on the averaged cross-section
\be
\int_{4}^{s_{max}} ds \sqrt{\frac{s}{s-4}}\sum_{\ell=4}^\infty (2\ell+1)a_\ell(s)<\lambda \frac{\pi(s_{max}-4)^4 (3 s_{max}-4)^3}{360(9 s_{max}^2-30 s_{max}+32)}\,.
\ee
This bound is valid for {\it any} $s_{max}>4$. Since this was reached using only the $m=2,n=1$ condition, stronger results are certainly possible. A generalization of eq.(\ref{ineqq}) is:
\be
\int_4^{s_{max}} \frac{ds}{s-\frac{4}{3}}\sqrt{\frac{s}{s-4}} \sum_{\ell=2m}^{\infty}(2\ell+1)a_\ell(s) {\mathcal B}_{m,m-1}^{(\ell)}(s)<\frac{(4m-3)4^{m-1}\Gamma(2m-\frac{3}{2})}{\pi \Gamma(m)\Gamma(m-\frac{1}{2})}\int_4^\infty ds \sqrt{\frac{s}{s-4}} \frac{a_{2m-2}(s)}{(s-4)^{2m-2}(s-\frac{4}{3})^{m+1}}\,.
\ee
We leave it to the reader to use this equation in imaginative ways!

\section{Feynman block expansion}{
We will now consider the Feynman block expansion of the four dilaton type II superstring tree amplitude after subtracting out the massless pole $-\frac{1}{s_1 s_2 (s_1+s_2)}$. Here $\a_0=2 \zeta (3)$. The s-channel  discontinuity is given by
$
\mathcal{A}(s_1;s_2)=\sum_{k=1}^{\infty}\frac{(-1)^{k+1} \Gamma \left(-s_2\right) \Gamma \left(k+s_2\right)}{(k!)^2 \Gamma \left(s_2+1\right) \Gamma \left(-k-s_2+1\right)}\pi \d(s_1-k)\,.
$ For demonstration purpose, we shall work in four spacetime dimensions, i.e., $\a=1/2$. The partial wave coefficients are given by
\be
\Phi(\s;1/2)a_\ell(\s)=\sum_{k=1}^{\infty}\pi \d(\s-k)~\frac{1}{2}\int_{-1}^{1}dx \frac{(-1)^{k+1} \Gamma \left(-\frac{1}{2} k (x-1)\right) \Gamma \left(\frac{1}{2} k (x+1)\right)}{(k!)^2 \Gamma \left(\frac{1}{2} k (x-1)+1\right) \Gamma \left(1-\frac{1}{2} k (x+1)\right)} C_{\ell}^{(1/2 )}(x)
\ee
Now, we can do the $\s$ integral by picking up $\s=k$ by $\d-$function. Using the results in the main text, we get the Feynman block expansion
\be
\label{eq:Feynman_string}
\begin{split}
\mathcal{M}(s_1, s_2)=\alpha_{0}+\sum_{\ell=0}^{\infty}\sum_{k=1}^{\infty} \frac{\left(2\ell+1\right)}{\left(k\right)^\ell}M_\ell(k;s_1,s_2)~\left[\frac{1}{2}\int_{-1}^{1}dx \frac{(-1)^{k+1} \Gamma \left(-\frac{1}{2} k (x-1)\right) \Gamma \left(\frac{1}{2} k (x+1)\right)}{(k!)^2 \Gamma \left(\frac{1}{2} k (x-1)+1\right) \Gamma \left(1-\frac{1}{2} k (x+1)\right)} C_{\ell}^{(1/2 )}(x)\right]\,,
\end{split}
\ee
For numerical purpose, we truncate the $k,\ell$ sum upto $k_{max}=6, L_{max}=6$. Numerical comparison is shown in the table \eqref{tab:string_feyndys}.
\begin{table}[hb]
\begin{tabular}{|c|c|c|c|c|}
\hline
 $s_1$ & $s_2$ & \text{Exact} & \text{Feynman} \\
\hline
 $\frac{1}{13}$ & $\frac{1}{10}$ & 2.45013 & 2.45012  \\
\hline
 $\frac{23}{83}$ & $-\frac{6}{17}$ & 2.61729 & 2.61724  \\
\hline
 $\frac{3}{10}+\frac{7 i}{10}$ & $-\frac{1}{11}-\frac{3 i}{13}$ & 1.70102\, +0.274874 i & 1.70092\, +0.274681 i \\
\hline
 $\frac{3}{5}+\frac{31 i}{97}$ & $\frac{1}{5}+\frac{i}{2}$ & 2.2343\, +1.66456 i & 2.23507\, +1.66368 i  \\
\hline
\end{tabular}
\caption{Numerical comparison with $k_{max}=6,L_{max}=6$}
\label{tab:string_feyndys}
\end{table}
}

{The convergence with $L_{max}$ is presented in table \eqref{tab:string_feyndys_conv} for $k_{max}=14$, from which we see that convergence is very good.  $a\approx 0.05, 0.08+0.002i, 0.36+0.24i$ respectively for the three points and shows that even away from $a=0$, convergence is very good. A more direct test of convergence would need a closed form expression for the contact terms for arbitrary spins, which we currently do not have. We leave the determination of the full range of $a$ where convergence holds for the Feynman block expansion for future work.}
\begin{table}[ht]
\begin{tabular}{|c|c|c|c|c|c|}
\hline
 $s_1$ & $s_2$ & \text{Exact} & $L_{\max }\text{=6}$ & $L_{\max }\text{=8}$ & $L_{\max }\text{=10}$ \\
\hline
 $\frac{1}{13}$ & $\frac{1}{13}$ & 2.43886  & 2.43886 & 2.43886 & 2.43886 \\
\hline
$ \frac{1}{13}+\frac{i}{3}$ & $\frac{1}{13}$ & 2.23803\, +0.12108 i & 2.23806\, +0.12105 i & 2.23804\, +0.12107 i & 2.23804\, +0.12107 i \\
\hline
 $\frac{1}{3}+\frac{i}{3}$ & $\frac{11}{13}$ & 7.66873\, +0.64394 i  & 7.66817\, +0.64304 i & 7.66853\, +0.64362 i & 7.66864\, +0.64380 i \\
\hline
\end{tabular}
\caption{The convergence with $L_{max}$ for $k_{max}=14$. We have exhibited up to 5 decimal places.}
\label{tab:string_feyndys_conv}
\end{table}

\section{Extremal solution to eq (8)}
{We define $\frac{\mathcal{W}_{p,q}}{\mathcal{W}_{p+q,0}}=\tilde{{W}}_{p,q}$. Very interestingly, we find that eq (8) (in main text) gives the following table with $\mu=0,\d_0=1$, where we compare with the closed string amplitude's $\tilde{{W}}_{p,q}^{(cl)}$\footnote{We remind the reader that this is just the massless pole subtracted amplitude used in the main text.}.
\be\nonumber
\begin{split}
&\begin{array}{|l|}
\hline
 \tilde{W}_{0,1}= -1.5 \\
\hline
 \tilde{W}_{0,1}^{\text{(cl)}}= -1.39348 \\
\hline
\end{array}~~\begin{array}{|l|l|}
\hline
 \tilde{W}_{1,1}=-2.5 & \tilde{W}_{0,2}= 1.5 \\
\hline
 \tilde{W}_{1,1}^{\text{(cl)}}= -2.47225 & \tilde{W}_{0,2}^{\text{(cl)}}= 1.47958 \\
\hline
\end{array}~~\begin{array}{|l|l|l|}
\hline
 \tilde{W}_{2,1}= -3.5 & \tilde{W}_{1,2}= 4 & \tilde{W}_{0,3}= -1.5 \\
\hline
 \tilde{W}_{2,1}^{\text{(cl)}}= -3.49239 & \tilde{W}_{1,2}^{\text{(cl)}}= 3.98908 & \tilde{W}_{0,3}^{\text{(cl)}}= -1.49593 \\
\hline
\end{array}\\
&\begin{array}{|l|l|l|l|l|l|}
\hline
 \tilde{W}_{5,1}= -6.5 & \tilde{W}_{4,2}= 17.5 & \tilde{W}_{3,3}= -25 & \tilde{W}_{2,4}= 20 & \tilde{W}_{1,5}= -8.5 & \tilde{W}_{0,6}= 1.5 \\
\hline
 \tilde{W}_{5,1}^{\text{(cl)}}= -6.49984 & \tilde{W}_{4,2}^{\text{(cl)}}= 17.4994 & \tilde{W}_{3,3}^{\text{(cl)}}= -24.9991 & \tilde{W}_{2,4}^{\text{(cl)}}= 19.9993 & \tilde{W}_{1,5}^{\text{(cl)}}= -8.49972 & \tilde{W}_{0,6}^{\text{(cl)}}= 1.49995 \\
\hline
\end{array}\\
&\begin{array}{|l|l|l|l|l|l|l|l|l|}
\hline
 \tilde{W}_{8,1}= -9.5 & \tilde{W}_{7,2}=40 & \tilde{W}_{6,3}= -98 & \tilde{W}_{5,4}= 154 & \tilde{W}_{4,5}= -161 & \tilde{W}_{3,6}= 112 & \tilde{W}_{2,7}= -50 & \tilde{W}_{1,8}= 13 & \tilde{W}_{0,9}= -1.5 \\
\hline
 \tilde{W}_{8,1}^{\text{(cl)}}= -9.5 & \tilde{W}_{7,2}^{\text{(cl)}}= 40 & \tilde{W}_{6,3}^{\text{(cl)}}=-98 & \tilde{W}_{5,4}^{\text{(cl)}}= 154 & \tilde{W}_{4,5}^{\text{(cl)}}= -161 & \tilde{W}_{3,6}^{\text{(cl)}}= 112 & \tilde{W}_{2,7}^{\text{(cl)}}= -50 & \tilde{W}_{1,8}^{\text{(cl)}}= 13 & \tilde{W}_{0,9}^{\text{(cl)}}= -1.5 \\
\hline
\end{array}
\end{split}
\ee
One can see the agreement is excellent, especially for larger $p$'s. Hence we conclude that low energy expansion of the closed string tree amplitude is very close to the extremal solution to eq (8) (in main text). Some related observations were made in some special cases in \cite{pedro,ras} recently. Note that in the above table using eq (8) (in main text) one can replace $\mathcal{W}_{p+q,0}$ by $\mathcal{W}_{1,0}$. In fact, a closed form expression for the extremal solution can be found if we assume $\mathcal{W}_{p,0}=\mathcal{W}_{1,0}=2$. This is given by
\be
\sum_{n,m=0}^\infty (-1)^n \frac{(2m+3n)\Gamma(m+n)}{ m! n!}x^m y^n=\frac{3y-2x}{(x-y-1)}=\left(\frac{s_1}{1-s_1}+\frac{s_2}{1-s_2}+\frac{s_3}{1-s_3}\right)\,.
\ee
This simply has the massive pole at $s_i=1$ with a constant subtraction. Quite remarkably this is a good approximation to the low energy expansion of the massless pole subtracted dilaton amplitude and extremizes eq(8). Further this provides a systematic procedure, in principle, to subtract out this extremal solution from the low energy expansion and go on to locating the 2nd massive pole. We have checked that the location of the pole comes out correctly at $s_i=2$ using the inequalities eq(8). 
}


\begin{thebibliography}{99}
\bibitem{foot1}
{ The enlarged domain of $a$ is worked out in \cite{AK}.}
\bibitem{foot3}
{A closed form can be found in terms of a ${}_4F_3$ hypergeometric function.}
\bibitem{exp1}Explicitly, for the first few cases, we have
$
\chi^{(0,1)}_n(\m,\d)=\frac{6 n+3}{6 \delta +4 \mu },~\chi^{(0,2)}_n(\m,\d)=\frac{9 (2 n (n+2)+3)}{4 (3 \delta +2 \mu )^2},~\chi^{(1,2)}_n(\m,\d)=\frac{6 n+3}{6 \delta +4 \mu }
$.
\bibitem{foot5}
{{In the large $\ell$ limit, $C_{\ell }^{(\alpha )}(1)D^{(n,m)}_{\ell,\a}\sim \ell^{2m+2\a-1}$. Convergence of the sum in \eqref{eq:largesDell} implies $a_\ell \sim \frac{1}{\ell^{2m+2\a+2}},~\forall m$. Therefore, $a_\ell$ is ``superpolynomially suppressed" in $\ell$.}}
\bibitem{foot8}
{See \cite{Martin, HS} for conventions. We are truncating the spin sum to $L_{max}$, assuming higher spins are suppressed. }
\end{thebibliography}
\end{document}